\documentclass[aps,prd,twocolumn,nofootinbib,superscriptaddress,preprintnumbers]{revtex4-1}
\usepackage{amsfonts,amsmath,amssymb,amsthm,braket,nicefrac}
\usepackage{booktabs,array}
\usepackage[utf8]{inputenc}
\usepackage{float,graphicx,hhline}
\usepackage{algorithm,algpseudocode}
\usepackage{appendix}
\usepackage[dvipsnames]{xcolor}
\usepackage{mathtools}
\usepackage{multirow}
\usepackage{tikz-cd}
\usepackage{slashed}
\usepackage{soul}
\usepackage{pifont}
\usepackage[caption=false]{subfig}
\usepackage[export]{adjustbox}
\usepackage[normalem]{ulem}

\AtBeginDocument{
\heavyrulewidth=.08em
\lightrulewidth=.05em
\cmidrulewidth=.03em
\belowrulesep=.65ex
\belowbottomsep=0pt
\aboverulesep=.4ex
\abovetopsep=0pt
\cmidrulesep=\doublerulesep
\cmidrulekern=.5em
\defaultaddspace=.5em
}

\usepackage{hyperref}
\hypersetup{
    pdftitle={},
    colorlinks=true,     
    linkcolor=blue,      
    citecolor=blue,      
    filecolor=blue,      
    urlcolor=blue        
}
\usepackage{cleveref}

\newcommand{\ssection}[1]{\emph{#1.---}}

\DeclareMathAlphabet{\mathbbold}{U}{bbold}{m}{n}

\let\Re\undefined
\DeclareMathOperator{\Re}{Re}
\DeclareMathOperator{\IM}{Im}
\DeclareMathOperator{\Tr}{Tr}

\begin{document}

\title{Flow-based sampling in the lattice Schwinger model at criticality}

\newcommand{\getMITAffiliation}{\affiliation{Center for Theoretical Physics, Massachusetts Institute of Technology, Cambridge, MA 02139, USA}}
\newcommand{\getNYUAffiliation}{\affiliation{Center for Cosmology and Particle Physics, New York University, New York, NY 10003, USA}}
\newcommand{\getDMAffiliation}{\affiliation{DeepMind, London, UK}}
\newcommand{\getIAIFIAffiliation}{\affiliation{The NSF AI Institute for Artificial Intelligence and Fundamental Interactions}}

\author{Michael~S.~Albergo}
\getNYUAffiliation
\author{Denis~Boyda}
\affiliation{Argonne Leadership Computing Facility, Argonne National Laboratory, Lemont IL-60439, USA}
\getMITAffiliation
\getIAIFIAffiliation
\author{Kyle~Cranmer}
\getNYUAffiliation
\author{Daniel~C.~Hackett}
\getMITAffiliation
\getIAIFIAffiliation
\author{Gurtej~Kanwar}
\affiliation{Albert Einstein Center, Institute for Theoretical Physics, University of Bern, 3012 Bern, Switzerland}
\getMITAffiliation
\getIAIFIAffiliation
\author{S\'{e}bastien~Racani\`{e}re}
\getDMAffiliation
\author{Danilo~J.~Rezende}
\getDMAffiliation
\author{Fernando~Romero-L\'opez}
\getMITAffiliation
\getIAIFIAffiliation
\author{Phiala~E.~Shanahan}
\getMITAffiliation
\getIAIFIAffiliation
\author{Julian~M.~Urban}
\affiliation{Institut f\"ur Theoretische Physik, Universit\"at Heidelberg, Philosophenweg 16, 69120 Heidelberg, Germany}

\preprint{MIT-CTP/5409}

\begin{abstract}
Recent results suggest that flow-based algorithms may provide efficient sampling of field distributions for lattice field theory applications, such as studies of quantum chromodynamics and the Schwinger model.
In this work, we provide a numerical demonstration of robust flow-based sampling in the Schwinger model at the critical value of the fermion mass. In contrast, at the same parameters, conventional methods fail to sample all parts of configuration space, leading to severely underestimated uncertainties.
\end{abstract}
\maketitle

Many important physical systems across particle and condensed matter physics can be described in the language of quantum field theory (QFT).
Lattice field theory (LFT) is the only known systematically improvable approach to ab-initio calculations of QFTs in non-perturbative regimes, such as quantum chromodynamics (QCD) at low energies.
LFT is based on the path-integral formulation of QFT, discretized on a Euclidean spacetime lattice. Monte Carlo techniques render the high-dimensional discretized path integral tractable by recasting the problem as statistical sampling: the expectation value of some observable $\mathcal{O}$ can be computed as
\begin{equation}
    \langle\mathcal O\rangle =  \frac{1}{\mathcal Z}\int dU\, e^{-S_E(U)} ~ \mathcal{O}(U)  \simeq \frac{1}{N} \sum_{i=1}^{N} \mathcal O(U_i),
\end{equation}
where $\mathcal Z$ is the partition function,
$S_E$ is the Euclidean action,
and $\{U_i\}$ is a set of $N$ samples of the lattice field degrees of freedom distributed as $p(U) = \exp[-S_E(U)]/\mathcal{Z}$. Statistical uncertainties decrease as $1/\sqrt{N}$ as the estimate converges to the true value.

In theories such as QCD, for which exact sampling algorithms are not known, Markov Chain Monte Carlo (MCMC) techniques are typically used. 
However, field samples or ``configurations'' from MCMC are correlated, i.e., subsequently generated configurations are not statistically independent. Depending on the MCMC approach, these ``autocorrelations'' may grow as the system is tuned towards criticality~\cite{Schaefer:2010hu}, e.g.~to describe universal properties of condensed matter theories or access the continuum or large-$N_c$ limits of gauge theories~\cite{DelDebbio:2002xa,Ce:2016awn}. 
Autocorrelations may become especially severe if MCMC updates are unlikely to generate transitions between modes that are separated in configuration space. This effect, known as ``freezing,'' can prevent effective exploration of the distribution for any practical sample size, amounting to an in-practice violation of ergodicity---a necessary condition for the validity of MCMC.

\begin{figure}[t]
    \centering
    \includegraphics{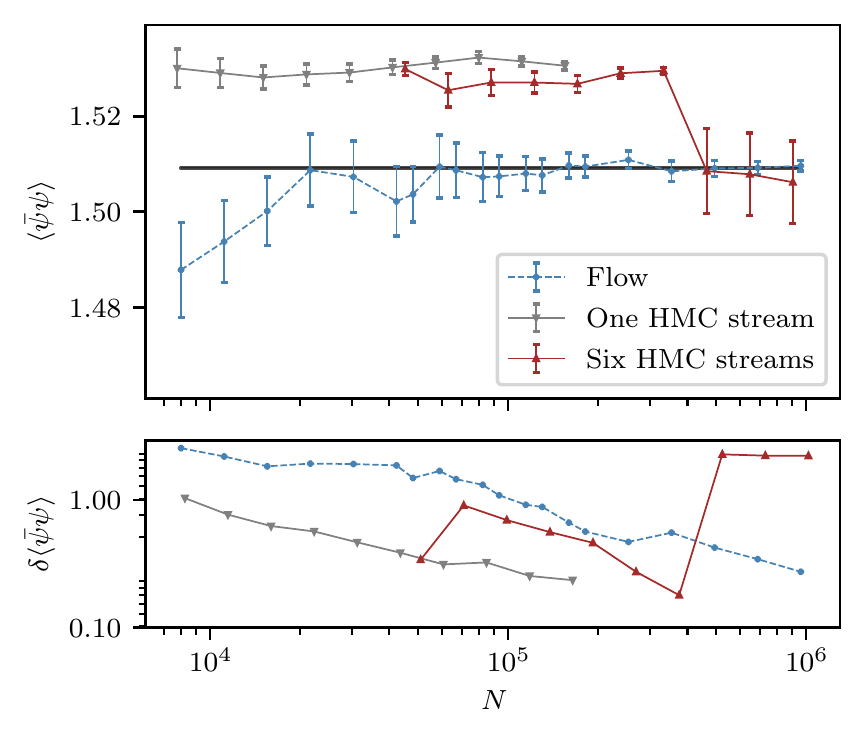}
    \vspace{-0.7cm}
    \caption{
    Demonstration of underestimated uncertainties when using HMC, in contrast to flow-based sampling which provides consistent results converging to a baseline value (black).
    The top panel shows estimates of the chiral condensate $\braket{\overline{\psi} \psi}$ in the Schwinger model at critical parameters $\beta=2.0$, $\kappa=0.276$, and $L=16$, as a function of the sample size $N$.
    The bottom panel shows the scaling of the statistical uncertainty $\delta \!\braket{\overline{\psi} \psi}$.
    Flow-based sampling (blue) converges to the baseline value with uncertainties scaling as $1/\sqrt{N}$.
    Meanwhile, HMC (gray, red) exhibits seemingly convergent uncertainties that are in fact severely underestimated, as indicated by the discrepancy with the baseline and sudden jumps when tunneling events occur (red).
    \vspace{-2em}
    }
    \label{fig:psibarpsi}
\end{figure}

Importantly, this affects Hybrid Monte Carlo (HMC), the state-of-the-art algorithm for sampling QCD field configurations, which generates samples by continuously evolving the fields through configuration space via Hamiltonian dynamics~\cite{DUANE1987216}.
These dynamics make the algorithm susceptible to freezing due to the topological properties of gauge fields, which divide the distribution into different modes or ``topological sectors''.
As the system is tuned towards criticality, increasingly large potential barriers suppress tunneling between sectors.
In contrast, emerging flow-based sampling algorithms~\cite{Albergo:2019eim,Kanwar:2020xzo,Boyda:2020hsi,Nicoli:2020njz,Albergo:2021bna,DelDebbio:2021qwf,Hackett:2021idh} propose random hops throughout configuration space, unaffected by density barriers.
While topological freezing presents a well-known obstacle to extending the reach of state-of-the-art lattice QCD calculations~\cite{Alles:1996vn,DelDebbio:2004xh,Schaefer:2010hu}, promising results of flow-based samplers in theories without fermions~\cite{Kanwar:2020xzo,Boyda:2020hsi,Foreman:2021ixr,Foreman:2021ljl}, without gauge fields~\cite{Albergo:2021bna}, or away from criticality~\cite{Finkenrath:2022ogg} suggest that these methods may provide a path towards mitigating freezing in this context.

In this Letter, we show that flow-based sampling can circumvent topological freezing in a fermionic gauge theory at criticality. Specifically, we provide a numerical demonstration in the Schwinger model (two-dimensional quantum electrodynamics) at the critical value of the fermion mass, illustrating that the flow-based approach is robust at sample sizes where HMC fails.
A stark example is shown in Fig.~\ref{fig:psibarpsi}, where HMC estimates of a key observable in the theory appear to be converging to a biased value, while in fact the uncertainty is underestimated due to insufficient sampling of the different topological sectors. In contrast, the flow-based sampling estimate is accurate with reliable uncertainties.

\ssection{Flow-based sampling for the Schwinger model}
Normalizing flow models~\cite{rezende2016variational,dinh2017density,JMLR:v22:19-1028} are based on applying a diffeomorphic ``flow'' transformation $f$ to (possibly high-dimensional) samples $z$ drawn from a  base distribution, $r(z)$.
This procedure yields samples ${U=f(z)}$ distributed according to the model density ${q(U) = r(z) \left|\det  \partial f / \partial z \right|^{-1}}$.
Flow-based sampling uses the model $q$ to approximate a target distribution $p$.
Neural networks can be used to construct an expressive and trainable flow, which can be optimized by minimizing the distance between $p$ and $q$.
Provably exact sampling that corrects for deviations between $p$ and $q$ can be obtained with independence Metropolis \cite{Metropolis:1953am,tierney1994markov} or reweighting; we use the former in this work.
These may be applied a posteriori, enabling embarrassingly parallel sampling that can provide practical advantages 
over HMC and sequential algorithms incorporating flows~\cite{Foreman:2021ixr,Foreman:2021ljl,Finkenrath:2022ogg}. 

Here, we apply flow-based sampling to the $N_f=2$ Schwinger model, a strongly interacting $(1+1)$d $U(1)$ gauge theory coupled to two fermions that exhibits similar features to QCD: confinement, spontaneous chiral symmetry breaking due to a chiral condensate, and nontrivial topology~\cite{Schwinger:1962tp,Coleman:1975pw}. It commonly serves as a toy model for QCD, and it has recently been used for testing new approaches to LFT~\cite{Dilger:1994ma,Albandea:2021lvl,Finkenrath:2022ogg,Hartung:2021wqg,Eichhorn:2021ccz}, including methods using quantum technologies~\cite{Funcke:2019zna,Butt:2019uul,Banuls:2019bmf}. It has also been used to study properties of QFTs~\cite{Coleman:1975pw,Coleman:1976uz,Smilga:1992hx,Giusti:2001xh,DAmico:2012wal,Shimizu:2014uva,Shimizu:2014fsa,Nagele:2018egu,Nagele:2020kef,Georgi:2020jik,Hu:2021gac}.

Wick rotating, discretizing, and integrating out the fermionic degrees of freedom yields a Euclidean lattice Schwinger model action amenable to Monte Carlo sampling~\cite{Gattringer:1997qc,Hip:1997em,Czaban:2013haa},
\begin{equation}\label{eq:action}
    S_E(U) = -\beta \sum_{x} \Re{P(x)} - \log \det D[U]^\dag D[U]\ ,
\end{equation}
given in terms of gauge links $U_{\mu}(x)$ at position~$x$ in direction~$\mu$.
The first term is the gauge action, where $\beta$ is the inverse of the squared gauge coupling, and the plaquette $P(x)$ is the smallest possible Wilson loop---a gauge-invariant product of links around a $1 \times 1$ square.
It is defined as
$P(x) = U_{0}(x) U_{1}(x+\hat{0}) U^\dagger_{0}(x+\hat{1}) U^\dagger_{1}(x),$
where $\hat{\mu}$ is the unit vector in the $\mu$ direction.
The second term, given in terms of the Wilson Dirac operator $D$~\cite{Wilson:1974sk,Wilson:1975id}, encodes the effect of fermions and gauge-fermion interactions.
The bare fermion mass $m_0$ is controlled by the hopping parameter $\kappa = 1/(4 + 2 m_0)$ that parametrizes~$D$. 

To achieve efficient sampling via a flow-based approach, it is critical to incorporate the physical properties of the target distribution.
For the Schwinger model specifically, gauge invariance imposes strong constraints on the target distribution, which we build into our models using the framework of gauge-equivariant flows on compact manifolds developed in Refs.~\cite{Kanwar:2020xzo,Boyda:2020hsi,Rezende:2020hrd}.
Another challenge is sampling of theories with fermionic degrees of freedom. Out of the four treatments in Ref.~\cite{Albergo:2021bna}, here we consider a ``marginal sampler'' using exact evaluation of the fermion determinant. This means that our model describes only gauge degrees of freedom, and we compute Eq.~\eqref{eq:action} exactly during training and for MCMC sampling.

Following Ref.~\cite{Kanwar:2020xzo}, gauge-equivariant flows are constructed by composing a sequence of equivariant coupling layers.
Each coupling layer updates an ``active'' subset of the links conditioned on a disjoint ``frozen'' subset. 
Different partitionings are used in each layer so that all variables are updated.
In each layer, gauge-invariant closed Wilson loops are computed from the frozen links and fed into a ``context function'' constructed from neural networks.
The outputs are used to parametrize the transformation of the active links, which is constrained to commute with gauge transformations.
Combined with a gauge-invariant base distribution, this yields a gauge-invariant model.

Unlike in the $\kappa=0$ limit of pure-gauge theory, the Schwinger model exhibits long-range correlations, with the correlation length defined by the inverse of the mass of the lightest particle.
At criticality, the correlation length diverges, which demands new architectural features over those employed in pure-gauge models~\cite{Kanwar:2020xzo}.
First, we use a subset of active links that is locally more sparse,
with each active link completely surrounded by frozen ones, to allow for better propagation of information.
Second, we provide larger $2 \times 1$ Wilson loops along with $1 \times 1$ plaquettes as inputs for context functions.
Third, our architecture includes dilated convolutions, which have translational equivariance and better context aggregation, i.e., an exponential expansion of the receptive field without loss of resolution or coverage~\cite{yu2015multi}.
Fourth, we parametrize our transformations using highly expressive neural splines \cite{durkan2019neural}.
Finally, we decay the learning rate over the course of training.

We train this flow model for the Schwinger model at criticality and compare the performance of flow-based MCMC using this model against that of HMC.
At finite lattice spacing, a diverging correlation length is realized by tuning $\kappa$ to its critical value, resulting in a vanishing renormalized fermion mass. To achieve this, we take $\beta=2.0$ and $\kappa=0.276$~\cite{Gattringer:1997qc} with a square lattice of extent $L=16$. 
Details of the architecture and training scheme are in the Supplementary Material.

\ssection{Advantages of flow-based sampling} 
A clear illustration of the advantages of flow-based sampling for the Schwinger model at criticality is given in Fig.~\ref{fig:psibarpsi}, which compares estimates of the chiral condensate from HMC with those from flow-based MCMC. This quantity,
\begin{equation}\label{eq:condensate}
    \langle \bar \psi \psi \rangle  = \frac{1}{V}\Tr D^{-1}[U],
\end{equation} 
is a simple fermionic observable whose value is correlated with the topological sectors and is therefore sensitive to freezing. We quantify uncertainties using the integrated autocorrelation time with the ``gamma method''~\cite{Wolff:2003sm} and compute the baseline result using an augmented version  of HMC that efficiently samples topological sectors.
Clearly, the single frozen HMC stream yields estimates that are manifestly inconsistent with the baseline result, indicating severely underestimated uncertainties even at very large sample sizes, $N \approx 10^5$. The dataset of samples from six independent HMC streams can incorporate information from multiple topological sectors even in the presence of freezing. However, as the figure shows, this estimate is still biased for $N \approx 10^5$ samples with incorrect uncertainties deceptively scaling as $1/\sqrt{N}$. 
The estimate becomes consistent with the ground truth only when $N \gtrsim 10^6$. The uncertainty, however, catastrophically increases---a clear indication of an ergodicity problem. This analysis suggests that affordable HMC stream lengths may not be sufficient to diagnose bias. By contrast, flow-based results converge smoothly to the baseline value, with errors scaling as $1/\sqrt{N}$.

\begin{figure}[t!]
    \centering
    \vspace{-0.1cm}
    \includegraphics{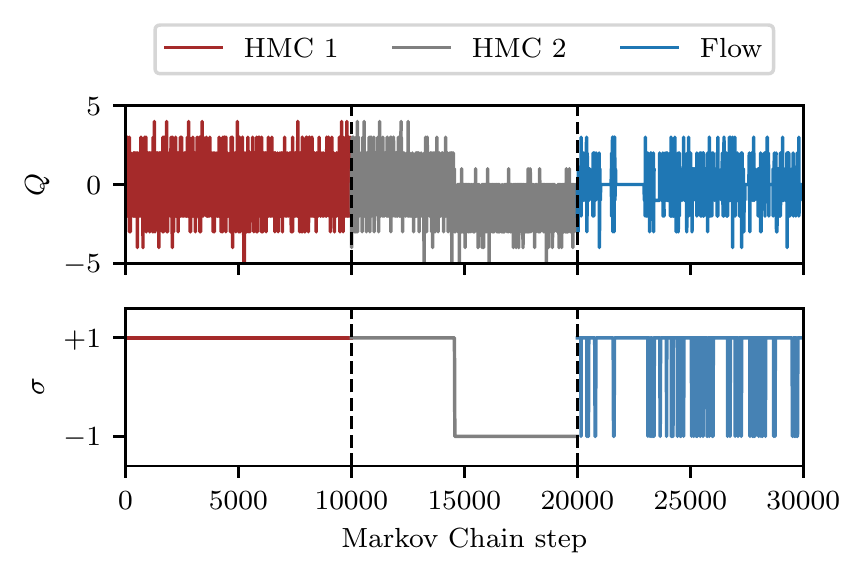}
    \vspace{-0.5cm}
    \caption{Monte Carlo history of the topological charge, $Q$~(top), and the sign of the real part of the determinant of the Dirac operator, $\sigma$~(bottom). Results shown are based on three different streams of configurations: two HMC streams (gray and green) and a stream from our flow model (blue). 
    In HMC, $Q$ exhibits ultraviolet fluctuations associated with discretization effects, but $\sigma$ rarely changes---an indication that true tunneling events between sectors are infrequent. Thus the HMC streams show clear evidence of topological freezing, while flow-based MCMC mixes rapidly.
    \vspace{-2em}
    }
    \label{fig:mixing}
\end{figure}

Figure~\ref{fig:mixing} provides a more direct illustration of freezing in the Monte Carlo histories of topological quantities. 
The topological sectors of the Schwinger model are distinguished by the integer-valued topological charge. A common discretization is~\cite{Shimizu:2014fsa}
\begin{equation}\label{eq:charge}
    Q = \frac{1}{2\pi} \sum_x  \IM \log P(x),
\end{equation}
where the imaginary part of the complex logarithm is restricted to $(-\pi,\pi]$.
Due to lattice artifacts, this observable fluctuates even when the topological sector is fixed.
A better-suited observable to identify true tunneling events was proposed in Ref.~\cite{Gattringer:1997qc}: 
\begin{equation}
    \sigma = \mathrm{sign}(\Re \, \det D).
\end{equation}
The value of $\sigma$ is positive (negative) for even (odd) topological sectors, and changes in its value are correlated with tunneling events.

Results for these observables based on two Markov chains generated with HMC (with different random number seeds) and one with flow-based sampling are illustrated in Fig.~\ref{fig:mixing}.
In the first HMC stream, $Q$ appears to fluctuate without any evidence of freezing. 
However, $\sigma$ is completely frozen for all samples shown, implying that these fluctuations arise from discretization effects and do not correspond to tunneling events between topological sectors.
In the second HMC stream, we see an abrupt change in the behavior of $Q$. This coincides with a change in $\sigma$, confirming that a true tunneling event has occurred.
By contrast, flow-based sampling exhibits rapid fluctuation in both $Q$ and $\sigma$, demonstrating sampling which rapidly mixes topological sectors.

A fair and comprehensive comparison of the costs of HMC and flow-based MCMC requires quantifying three factors for each: setup costs, the raw computational cost of a sampling step, and the sampling efficiency (i.e.,~the degree of autocorrelation).
Setup costs---predominantly, equilibration for HMC and training for flows---are particularly difficult to compare in this case.
Full equilibration of HMC requires observing and discarding many tunneling events, which occur stochastically, while training costs for the flow-based approach may vary over orders of magnitude depending on the training scheme.
Raw computational costs may be measured directly, but depend strongly on implementation details. On the same GPU hardware, we find that flow-based MCMC steps are $\sim 10$ times less expensive than HMC trajectories, due to the frequent inversions of the Dirac operator in HMC. However, there is room for optimization in both cases.

Nevertheless, disregarding setup and raw computational costs, an approximate comparison of sampling efficiency is sufficient to show the advantage of flow-based sampling over HMC.
Each algorithm exhibits some characteristic time between tunneling events; a chain with many times that number of steps will be required to incorporate information from all topological sectors with appropriate weights.
For these HMC parameters, we find tunneling events are separated by $\sim 20\mathrm{k}$ trajectories on average. 
Meanwhile, sampling with our flow model, the topological sector changes every $\sim 6$ steps on average.
Thus, for this model at the parameters investigated, we estimate that the advantage in sampling efficiency of flow-based MCMC over HMC is more than three orders of magnitude.

\ssection{Conclusion and outlook}
In this Letter, we demonstrate that flow-based sampling can be applied to lattice gauge theories with fermion content at criticality.
Specifically, we have developed an architecture that can successfully model long-range correlations in the Schwinger model at vanishing renormalized fermion mass. 
The resulting flow-based sampler does not suffer from topological freezing at these parameters and thus outperforms HMC by orders of magnitude.
These results represent an important milestone in first-principles calculations in gauge field theories with fermions, such as QCD, using provably exact machine learning.

Importantly, the flow models described here may serve as an ``engine'' for a much broader class of sampling algorithms.
For example, flow-based MCMC updates may be interleaved with steps of HMC~\cite{Hackett:2021idh} or other MCMC algorithms~\cite{Gabrie:2021tlu}.
Such composite algorithms may provide improved sampling over either method alone.
Another possible improvement is using the flow models developed here inside a hierarchical multilevel MCMC scheme, as proposed by Ref.~\cite{Finkenrath:2022ogg}.
Finally, our technical advances can be adapted for more general MCMC schemes, e.g.,~generalizations of the HMC algorithm~\cite{Foreman:2021ixr,Foreman:2021ljl} and stochastic normalizing flows~\cite{Matthews:2022sds, wu2020stochastic, nielsen2020survae}. 

Challenges remain on the road to large-scale applications, such as state-of-the-art QCD calculations.
The sampling approach here employs exact evaluation of the fermion determinant, but more scalable approaches will be needed for larger volumes and theories in higher dimensions; extending the machine-learned stochastic determinant estimators of Ref.~\cite{Albergo:2021bna} to lattice gauge theories presents a promising opportunity.
If the success demonstrated here for the Schwinger model can be extended to other theories, and in particular at scale, it will have widespread impact across nuclear and particle physics, as well as in condensed matter applications.

\newpage
\ssection{Acknowledgements}
We thank Ryan Abbott for useful comments on this manuscript. GK, DCH, FRL, and PES are supported in part by the U.S.\ Department of Energy, Office of Science, Office of Nuclear Physics, under grant Contract Number DE-SC0011090. PES is additionally supported by the National Science Foundation under EAGER grant 2035015, by the U.S.\ DOE Early Career Award DE-SC0021006, by a NEC research award, and by the Carl G and Shirley Sontheimer Research Fund. GK is additionally supported by the Schweizerischer Nationalfonds. KC and MSA are supported by the National Science Foundation under the award PHY-2141336. MSA thanks the Flatiron Institute for their hospitality. DB is supported by the Argonne Leadership Computing Facility, which is a U.S. Department of Energy Office of Science User Facility operated under contract DE-AC02-06CH11357. This work is supported by the Deutsche Forschungsgemeinschaft (DFG, German Research Foundation) under Germany's Excellence Strategy EXC 2181/1 - 390900948 (the Heidelberg STRUCTURES Excellence Cluster), the Collaborative Research Centre SFB 1225 (ISOQUANT), and the U.S.\ National Science Foundation under Cooperative Agreement PHY-2019786 (The NSF AI Institute for Artificial Intelligence and Fundamental Interactions, \url{http://iaifi.org/}). This work is associated with an ALCF Aurora Early Science Program project, and used resources of the Argonne Leadership Computing Facility, which is a DOE Office of Science User Facility supported under Contract DEAC02-06CH11357. The authors acknowledge the MIT SuperCloud and Lincoln Laboratory Supercomputing Center~\cite{reuther2018interactive} for providing HPC resources that have contributed to the research results reported within this paper. Numerical experiments and data analysis used PyTorch~\cite{NEURIPS2019_9015}, Horovod~\cite{sergeev2018horovod}, NumPy~\cite{harris2020array}, and SciPy~\cite{2020SciPy-NMeth}.
Figures were produced using matplotlib~\cite{Hunter:2007}.

\bibliographystyle{utphys}
\bibliography{main}

\clearpage
\widetext
\begin{center}
\textbf{\large Supplementary Material}
\end{center}
\setcounter{equation}{0}
\setcounter{figure}{0}
\setcounter{table}{0}
\setcounter{page}{1}
\makeatletter
\renewcommand{\theequation}{S\arabic{equation}}
\renewcommand{\thefigure}{S\arabic{figure}}

\section{Fermionic action}

The fermion part of the Schwinger model action for $N_f$ degenerate flavors can be written as
\begin{equation}
    S_f(U,\psi,\bar{\psi}) = \sum_{f=1}^{N_f} \sum_{x,y} \bar{\psi}_f^{\beta}(y) D[U](y,x)^{\beta \alpha} \psi_f^{\alpha}(x) \ ,
\end{equation}
where $\psi_f^{\alpha}(x)$ denotes a two-component fermion field with flavor $f$ and spin index $\alpha \in \{1,2\}$.
We employ the Wilson discretization~\cite{Wilson:1974sk,Wilson:1975id} of the lattice Dirac operator $D[U]$, given by
\begin{equation}\label{eq:dirac-wilson}
    D[U](y, x)^{\beta \alpha} = \delta(y-x) \delta^{\beta \alpha} 
    - \kappa \sum_{\mu=0,1} \Big\{  [1-\sigma_{\mu}]^{\beta \alpha} U_{\mu}(y) \delta(y - x + \hat{\mu}) 
    +  [1+\sigma_{\mu}]^{\beta \alpha} U^\dagger_{\mu}(y-\hat{\mu}) \delta(y - x - \hat{\mu}) \Big\}\ ,
\end{equation}
where $\sigma_{\mu}=(\sigma_x,\sigma_y)$, with $\sigma_{x,y}$ denoting the usual Pauli matrices, and $\hat{\mu}$ a unit vector in direction $\mu$.
The $\delta$ functions are defined incorporating anti-periodic boundary conditions in the time direction.

Integrating out the fermionic Grassmann variables with $N_f=2$ allows an evaluation of the action in terms of the determinant of $D$---see Eq.~\eqref{eq:action} in the main text. 
Using matrix notation for the left and right spatial and spinor indices of $D$ (i.e.,~$y,\beta$ and $x,\alpha$), the fermionic action reads
\begin{equation}
    S_f(U) = - \log \det D[U]^\dag D[U]\ .
\end{equation}
This version of the action, without additional pseudofermionic degrees of freedom, is used throughout this work. This includes for the definition of the loss in training flow-based models---see Eq.~\eqref{eq:loss} below---and in sampling for both HMC and flow-based MCMC.

\section{Flow architecture and training scheme}

The base distribution $r(U)$ for the flow model described in the main text is an independent uniform distribution over the $U(1)$ Haar measure on each link (i.e.,~$A$ uniform over $[0,2\pi)$ for $U=e^{iA}$).
We sample this distribution using the NumPy uniform random number generator to circumvent precision issues with the implementation in PyTorch~1.9.
Each gauge-equivariant coupling layer~\cite{Kanwar:2020xzo} updates the active subset of the links, 
\begin{equation}
    M_{\mu\nu}^k = \left\{ U_{\mu}\big( (4n+k)\hat{\mu} + 2m \hat{\nu} \big) \big| ~ \forall\, n,m \in \mathbb{Z}  \right\} 
    \cup
    \left\{ U_{\mu}\big(   (4n+2+k)\hat{\mu} + (2m+1)\hat{\nu} \big) \big| ~ \forall\, n,m \in \mathbb{Z}  \right\},
\end{equation}
where $k,\mu,\nu$ change for each layer.
The flow is constructed by iterating through $k\in\{0,1,2,3\}$ in order first with $\mu=0,\nu=1$, then for $\mu=1,\nu=0$, such that all links are updated in 8 layers.
We use a model with 48 layers, so each link is updated 6 times.
For active loops, we use the plaquettes that project forwards from their corresponding active links (``active plaquettes'').

Each layer has its own convolutional neural network that outputs the parameters defining the transformation of active plaquettes.
Each neural network takes six input channels:
\begin{equation}
    \cos \theta_P,\ \sin \theta_P,\ \cos \theta_{2\times1} ,\ \sin \theta_{2\times1},\ \cos \theta_{1\times2} ,\ \sin \theta_{1\times2},
\end{equation}
where $\theta_{P}$ is the argument of the plaquette, and $\theta_{2\times1}$($\theta_{1\times2}$) is the argument of the $2\times1$ ($1\times2$) Wilson loop. The $\sin$ and $\cos$ are applied to ensure that the input is a continuous function of the gauge fields.
Each neural network is built from three convolutions with kernel size 3 and dilation factors $1,2,3$ in order (where 1 is a standard undilated convolution).
Between each intermediate convolution, there are 64 hidden channels, and the final output has 10 channels. 
After each intermediate convolution we use LeakyReLU activations, but no activation is applied to the final output. 
The 10 output channels are used to parametrize the positions and slopes of the 3 knots of a circular rational quadratic spline $s(\theta_{P})$, as well as an overall offset $t$. 
These are used to transform the active plaquettes $\theta_{PA}$ as $\theta_{PA}' = s(\theta_{PA}) + t$.
The active links are updated by inferring a link transformation that will induce the transformation of the active loops.
The total number of parameters of the model is $\sim 2.2 \times 10^6$.

We initialize the model weights using the PyTorch 1.9 defaults.
We use a self-training scheme where the loss function is a stochastic estimate of the Kullback-Leibler divergence~\cite{Kullback:1951} made with $q$-distributed samples generated by the model,
\begin{equation}
    D_{\mathrm{KL}}(q||p) 
    = \int dU \, q(U) \log \frac{q(U)}{p(U)} 
    \approx \frac{1}{B} \sum_{i=1}^B \left[ \log q(U_i) + S(U_i) \right] + (\text{const}), \quad (U_i \sim q), \label{eq:loss}
\end{equation}
where $B=3072$ is the batch size (i.e.,~number of field samples generated for each gradient step), and the constant does not affect optimization and so is not computed.
We train using the Adam optimizer with standard PyTorch 1.9 parameters.
We decay the learning rate every 30k gradient updates following the schedule $ [ 3 ,\ 1.5,\ 0.75,\ 0.4,\ 0.2,\ 0.1,\ 0.05 ]\cdot 10^{-4}$.
In total, we train the model for 210k gradient steps.
For each step, we clip the absolute value of each parameter gradient to be less than or equal to 0.1, and the norm over all gradients to be less than or equal to 10.
We use a mixed-precision approach for both training and sampling, using single precision for the neural network evaluations but double precision for all other computations.

We construct an ensemble from the flow model by drawing independent samples as proposals for independence Metropolis to ensure exactness. The Metropolis acceptance rate is $\sim 17\%$.
To obtain a sample of $N$ configurations, we take the initial $1.25N$ configurations in the chain and discard the first $0.25N$ for equilibration.

\section{HMC Details}

The HMC results presented in this work are for HMC applied to the exact determinant action Eq.~\eqref{eq:action}, without using any additional pseudofermionic degrees of freedom.
That is, the action for accept/reject tests is computed directly using Eq.~\ref{eq:action}, and molecular dynamics forces are computed using exact derivatives with respect to $S_f[U]$. We use trajectories of length $\tau_\text{HMC}=1$  divided into 10 steps, yielding an acceptance rate of $94\%$. 
We use double precision.

The HMC data used in Fig.~\ref{fig:psibarpsi} (and Fig.~\ref{fig:topsus} below) are taken from streams $2 \times 10^5$ trajectories long, with no trajectories discarded between measurements.
Each stream is initialized from a hot start, i.e.,~all links drawn from independent uniform distributions.
We use the same scheme for equilibration cuts as described above for flows.
For the six-stream example, an equal number of configurations is taken from each stream.

For the Schwinger model, it is possible to implement an augmentation step for HMC that proposes hops to other topological sectors~\cite{Albandea:2021lvl}, i.e.,\ configurations $U'$ such that $0 \neq Q(U') - Q(U) \equiv \varDelta Q \in \mathbb{Z}$. This is achieved by distributing the proposed change across links according to
\begin{equation}
\begin{aligned}
    U_0'(x) &= \exp\left(-2\pi\mathrm{i} \frac{\varDelta Q}{V} x_1 \right)\, U_0(x) \\
    U_1'(x) &= \exp\left(2\pi\mathrm{i} \frac{\varDelta Q}{V} x_0 \delta_{x_1,L-1}\right)\, U_1(x) \ ,
\end{aligned}
\end{equation}
with coordinates in lattice units, i.e.,\ $x_i \in\{ 0, \dots, L-1\}$. 
For simplicity, we restrict ${\varDelta Q \in \{-2,-1,0,1,2\}}$ and propose each $\varDelta Q$ with equal probability. 
The proposal is accepted or rejected with a standard Metropolis step. The acceptance rate of these proposals is 34\% for these parameters.

Interleaving this augmentation with HMC steps produces an algorithm with different properties than HMC alone (augmented HMC), and no equivalent construction is known for many theories, including QCD. Thus we solely use this method to obtain baseline results for the chiral condensate in Fig.~\ref{fig:psibarpsi} and the topological susceptibility in Fig.~\ref{fig:topsus}. To do so, an ensemble of $1.2 \times 10^7$ configurations is produced using these augmentation hits alternated with HMC steps. The baseline results are estimated to be $\langle \bar \psi \psi \rangle = 1.50918(9)$, $\langle\chi_Q\rangle = 0.003875(4)$.

\begin{figure}[H]
    \centering
    \includegraphics{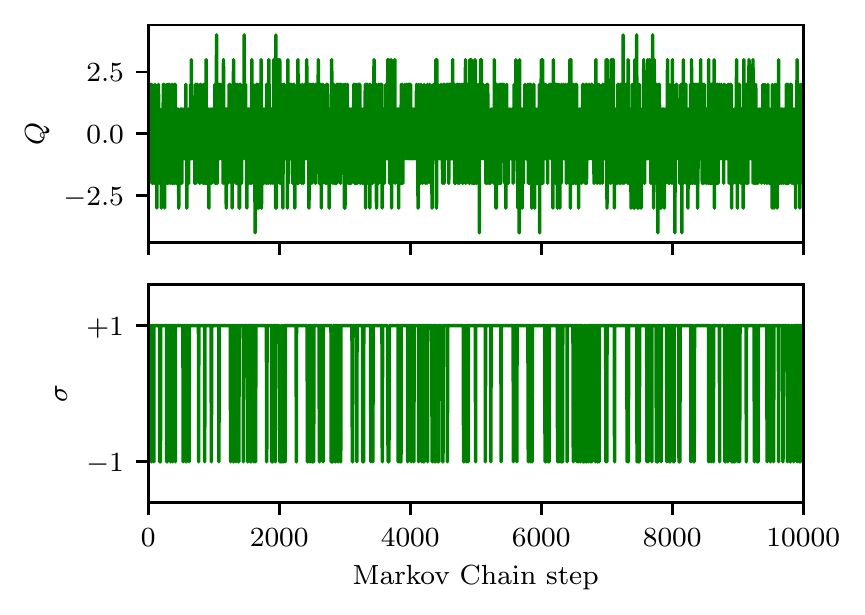} \\
    \caption{ 
    Similar to Fig.~\ref{fig:mixing}, but computed with augmented HMC.
    }
    \label{fig:augHMC}
\end{figure}

In Fig.~\ref{fig:augHMC} we show part of the Monte Carlo history of $Q$ and $\sigma$ for the baseline run with augmented HMC. As can be seen, rapid fluctuations occur at a scale comparable to the flow model. In particular, $\sigma$ exhibits an asymmetric distribution along positive and negative values, as expected from the higher total weight of even topological sectors.

\section{Topological susceptibility}

\begin{figure}[H]
    \centering
    \includegraphics{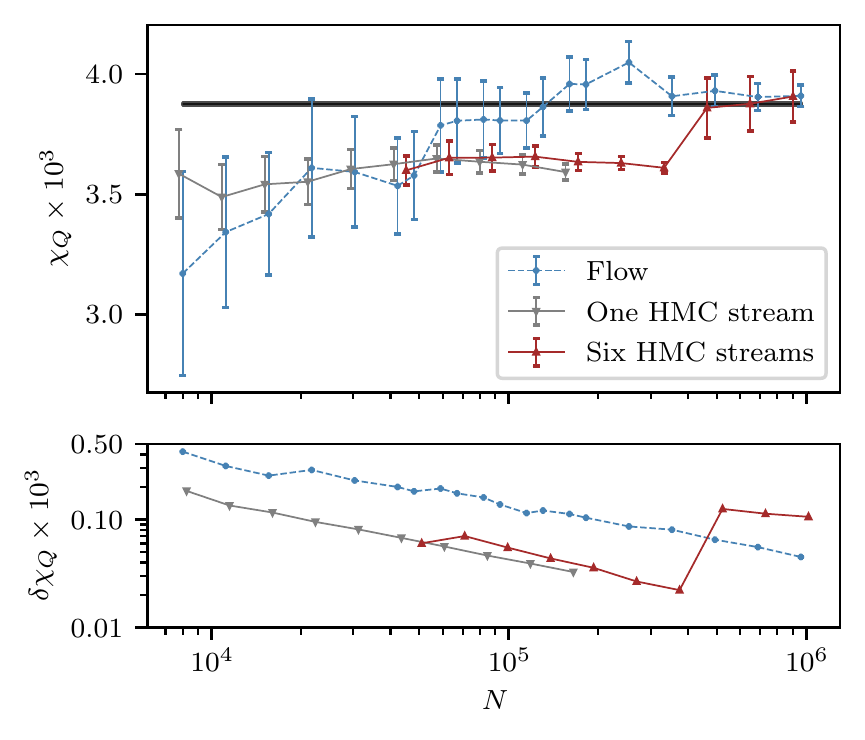} \\
    \caption{
    Similar to Fig.~\ref{fig:psibarpsi}, but for the topological susceptibility ${\chi_Q}$.
    }
    \label{fig:topsus}
\end{figure}

It is also interesting to look at the bias of an observable other than the chiral condensate illustrated in Fig.~\ref{fig:psibarpsi} in the main text. The topological susceptibility, defined as
\begin{equation}\label{eq:susceptibility}
    \chi_Q = \frac{1}{V} \langle Q^2 \rangle,
\end{equation}
is shown computed from both HMC and the flow-based approach in Fig.~\ref{fig:topsus}. We observe similar behavior in this observable as for the chiral condensate of Fig.~\ref{fig:psibarpsi}---an underestimation of uncertainties in the HMC results that is difficult to discern without very high statistics.

\end{document}